\documentclass[cits,a4paper]{PoS}
\usepackage{amsmath,amssymb}
\usepackage[tight,TABTOPCAP]{subfigure}
\usepackage{graphicx}
\usepackage{grffile}
\usepackage[font=small,labelfont=bf]{caption}

\def\mbar{\overline{m}}
\newcommand{\psibar}{\overline \psi}
\newcommand{\nft}{N_\mathrm{f}=2}

\graphicspath{{figures/}}

\title{Study of the Couplings of QED and QCD from the Adler Function}

\ShortTitle{Study of the couplings of QED and QCD from the Adler function}

\author{

  Anthony Francis${\,}^a$, \speaker{Gregorio Herdo\'iza}${\,}^b$, Hanno Horch${\,}^c$,
  Benjamin J\"{a}ger${\,}^{d}$, Harvey~B.~Meyer${\,}^{a,c}$, Hartmut Wittig${\,}^{a,c}$\\
  \llap{$^a$} Helmholtz Institute Mainz,\\
  Johannes Gutenberg-Universit{\"a}t, 55099 Mainz, Germany \\
  \llap{$^b$}Instituto de F\'isica Te\'orica UAM/CSIC and
  Departamento de F\'isica Te\'orica,\\
  Universidad Aut\'onoma de Madrid, Cantoblanco, E-28049
  Madrid, Spain\\
  \llap{$^c$} PRISMA Cluster of Excellence, Institut f{\"u}r Kernphysik,\\
  Johannes Gutenberg-Universit{\"a}t, 55099 Mainz, Germany\\
  \llap{$^d$} Department of Physics, Swansea University, Swansea, United Kingdom\\
  E-mails:
  \email{\{francis,horch,meyerh,wittig\}@kph.uni-mainz.de,
    gregorio.herdoiza@uam.es, B.Jaeger@swansea.ac.uk} 

}

\abstract{

  The contribution from hadronic vacuum polarisation effects is
  responsible for a large fraction of the theoretical uncertainty in the
  running of the QED coupling. The current level of uncertainty has
  become a limitation for electroweak precision tests. We use lattice
  QCD simulations with two flavours of O$(a)$ improved Wilson fermions
  to determine the Adler function in a broad range of the momentum transfer
  $Q^2$. The running of the QED coupling, including valence
  contributions from $u$, $d$, $s$ and $c$ quarks, is compared to
  phenomenological results at intermediate $Q^2$ values. In the large
  $Q^2$ regime, the lattice determination of the Adler function is
  fitted to perturbation theory in order to examine the feasibility of a
  determination of the strong coupling constant.

}

\FullConference{The 32nd International Symposium on Lattice Field Theory,\\
  23-28 June, 2014\\
  Columbia University New York, NY}

\begin{document}


\section{Introduction}

The charge renormalisation in QED is due to the energy dependence
induced by photon vacuum polarisation effects. The running of the QED
coupling $ \alpha(Q^2)$ can be written as follows,
\begin{equation}\label{eq:aldef}
  \alpha(Q^2)=\frac{\alpha}{1- \Delta
    \alpha_{\rm QED}(Q^2)} \,,
\end{equation}
where $\alpha$ is the fine structure constant providing the classical
charge normalisation at vanishing photon virtuality, $Q^2=0$. $\Delta
\alpha_{\rm QED}(Q^2)$ is related to the subtracted vacuum
polarisation function (VPF). The relative size of the contributions to
$\Delta \alpha_{\rm QED}(Q^2)$ from loops of quarks and leptons is
comparable. Recent determinations of $\alpha(Q^2)$ at
$Q^2=0$~\cite{Agashe:2014kda} and at the $Z$-boson
mass~\cite{Davier:2010nc} read,
\begin{align}
  \alpha &= 1/137.035 999 074(44)~~~[0.3\,{\rm ppb}]\,,\\
  \alpha(M_Z^2) &=
  1/128.952(14) \hspace*{1.45cm} [10^{-4}]\,,
\end{align}
where the values in brackets denote the relative error. While the
central values shift by merely $\sim 6\%$, the errors instead
deteriorate by five orders of magnitude when running up to the
$Z$-pole. Hadronic effects are responsible for a large fraction of
this loss of precision. A phenomenological
approach~\cite{Davier:2010nc,Jegerlehner:2011mw,Hagiwara:2011afg}
using dispersive methods together with experimental measurements of
the cross-section, $\sigma(e^+e^- \to {\rm hadrons})$, is at present
used to measure the hadronic contribution to the running of the QED
coupling, $\Delta \alpha_{\rm QED}^{\rm had}$. Lattice QCD provides an
alternative method to determine this observable.

Contrary to the case of the lowest-order hadronic contribution to the
anomalous magnetic moment of the muon, $a_\mu^{\rm HLO}$, which is
dominated by the low-$Q^2$ regime, $\Delta \alpha_{\rm QED}^{\rm had}$
receives sizeable contributions from all energy regions. In order to
run $\Delta \alpha_{\rm QED}^{\rm had}$ to the $Z$-boson mass, a
matching to perturbative QCD (pQCD) is often done at $Q_{\rm match}^2
\approx 6\,{\rm GeV}^2$. At this energy scale, the dispersive approach
tends to be poorly constrained by cross-section measurements in the
region of $s$ between $1$ to $4\,{\rm
  GeV}^2$~\cite{Jegerlehner:2008rs}. An interesting question is
therefore to study whether there is a $Q^2$-interval where current
lattice QCD calculations could reach a comparable accuracy than the
dispersive approach. This would provide a valuable test in a context
where the present error on $\Delta \alpha_{\rm QED}$ has become a
limitation for electroweak precision tests.


\section{Lattice QCD Study of the Running of the QED Coupling}
\label{sec:alp}

The hadronic vacuum polarisation tensor, depending on Euclidean
momentum $Q$ is given by,
\begin{equation}\label{eq:pol_tensor}
  \Pi_{\mu\nu}(\hat Q)= \int d^4x
  \,e^{iQx} \,\langle J_\mu(x)J_\nu(0) \rangle \,,
\end{equation}
where, in practice, the lattice momentum $\hat{Q}_\mu={2}/{a}
\,\sin\left({aQ_\mu}/{2}\right)$ is used. The vector current
reads,
\begin{equation}
  J_\mu(x)=\sum_{{\rm f}=u,\,d,\,s,\,c,\dots} \, Q_{\rm f} \, \psibar_{\rm
    f}(x)\gamma_\mu \psi_{\rm f}(x) \,,
  \label{eq:current}
\end{equation}
where $Q_{\rm f}$ is the electric charge of the quark flavour ${\rm
  f}$. The VPF $\Pi(\hat Q^2)$ is extracted from,
\begin{equation}
  \Pi_{\mu\nu}(\hat Q)=(\hat Q_\mu \hat Q_\nu - \delta_{\mu\nu} \hat Q^2)
  \,\Pi(\hat Q^2)\,.
  \label{eq:vpcont}
\end{equation}
The subtracted VPF, $\widehat{\Pi}(\hat Q^2) = \Pi(\hat Q^2)-\Pi(0)$,
is directly related to the hadronic contribution to the running of the
electromagnetic coupling, $\Delta\alpha_{\rm QED}^{\rm had}(\hat Q^2)
= 4 \pi \alpha \, \widehat{\Pi}(\hat Q^2)$\,. A related physical
quantity that also allows to determine $\Delta\alpha_{\rm QED}^{\rm
  had}$ is the Adler function, defined in the following way,
\begin{equation}
  D(\hat Q^2) \ = \ 12\pi^2
  \,\hat Q^2\,\frac{d\,\Pi\,(\hat Q^2)}{d\hat Q^2}
  \ =\ \frac{3\pi}{\alpha}\,\hat Q^2\,\frac{d}{d\hat Q^2}\Delta
  \alpha_{\rm QED}^{\mathrm{had}}(\hat Q^2)\,.
  \label{eq:adler}
\end{equation}
The Adler function has a smooth dependence on $\hat Q^2$ due to the
absence of resonance effects in the space-like domain. It is therefore
a useful quantity to examine the $\hat Q^2$-regime where pQCD applies.

\begin{figure}[!t]
  \begin{minipage}{\textwidth}
    \begin{minipage}[b]{0.49\textwidth}
      \hspace*{-0.3cm}
      \centering
      \begin{tabular}{cccccr}
        \hline
        Ens. & $a\,[\mathrm{fm}]$ & $V/a^4$ & $M_{\rm PS}$ & $M_{\rm PS} L$
        & $N_{\rm meas}$\\
        \hline
        A3 & $0.079$ & $64 \times 32^3$  & $473$ & $6.0$ & $1004$\\
        A4 &         & $64 \times 32^3$  & $363$ & $4.7$ & $1600$\\
        A5 &         & $64 \times 32^3$  & $312$ & $4.0$ & $1004$\\
        B6 &         & $96 \times 48^3$  & $267$ & $5.1$ & $1224$\\
        \hline
        E5 & $0.063$ & $64 \times 32^3$  & $456$ & $4.7$ & $4000$\\
        F6 &         & $96 \times 48^3$  & $325$ & $5.0$ & $1200$\\
        F7 &         & $96 \times 48^3$  & $277$ & $4.2$ & $1000$\\
        G8 &         & $128 \times 64^3$ & $193$ & $4.0$ & $820$\\
        \hline
        N5 & $0.050$ & $96 \times 48^3$  & $430$ & $5.2$ & $1392$\\
        N6 &         & $96 \times 48^3$  & $340$ & $4.1$ & $2236$\\
        O7 &         & $128 \times 64^3$ & $261$ & $4.4$ & $552$\\
        \hline
      \end{tabular}
      \captionof{table}{Ensembles of O$(a)$ improved Wilson fermions
        used in this work. Approximate values of the lattice spacing
        $a$ and of the pion mass $M_{\rm PS}$ (in $\mathrm{MeV}$) as well
        as information about the lattice volume and the number of
        measurements $N_{\rm meas}$ are provided.}
      \label{tab:ens}
    \end{minipage}
    \hspace*{0.45cm}
    \begin{minipage}[b]{0.45\textwidth}
      \includegraphics[height=0.95\linewidth]{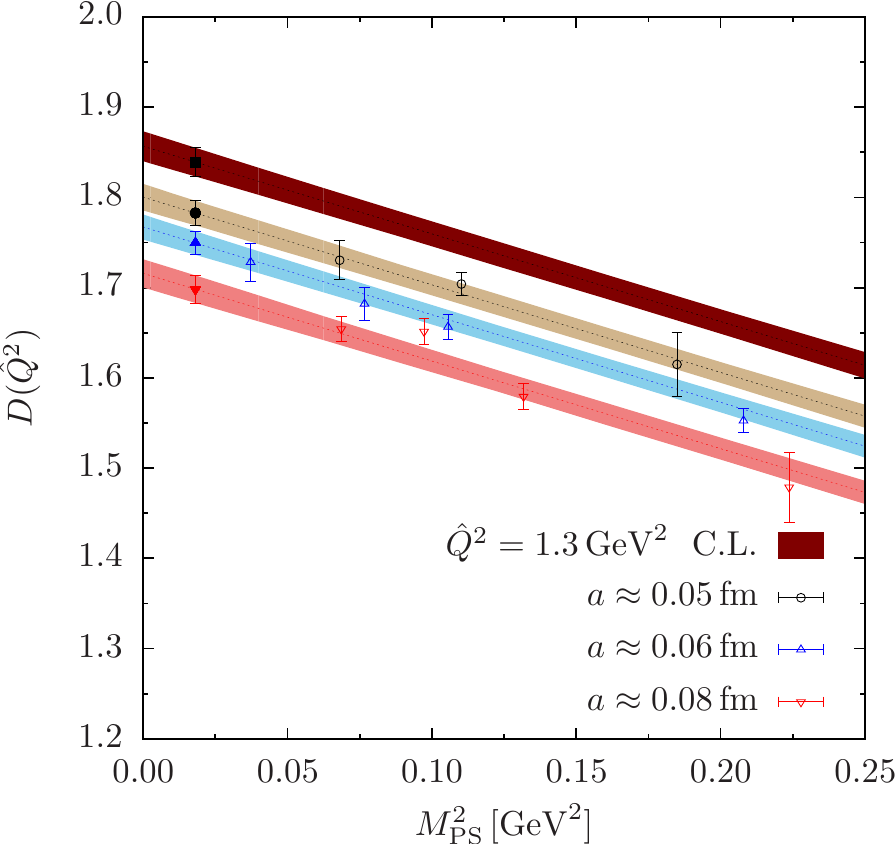}
      \captionof{figure}{Pion mass dependence of the Adler function at
        fixed $\hat Q^2=1.3\,{\rm GeV^2}$. The upper band, denoted by
        `C.L.', is the continuum limit estimate. The leftmost (filled)
        symbols indicate the extrapolated values at the physical pion
        mass.}
      \label{fig:DMpi}
    \end{minipage}
  \end{minipage}
\end{figure}

The calculation of $D(\hat Q^2)$ is based on a set of CLS lattice
ensembles (c.f. table~\ref{tab:ens}) with two flavours of O($a$)
improved Wilson fermions at three values of the lattice spacing and
pseudoscalar meson masses down to $190$\,MeV satisfying the condition,
$M_{\rm PS}\,L \geq 4$. The use of partially twisted boundary
conditions~\cite{DellaMorte:2011aa} increases the density of $\hat
Q^2$ values and allows to construct the Adler function from the
numerical derivative of the VPF. We refer to
refs.~\cite{Horch:2013lla,DellaMorte:2014rta,Francis:2014dta} for more
details about our procedure to determine $D(\hat Q^2)$. The lattice
data for $D(\hat Q^2)$ is parametrised by a fit ansatz that describes
simultaneously the $\hat Q^2$ dependence through Pad\'e approximants
and the continuum and chiral
extrapolations~\cite{DellaMorte:2014rta}. Once the parameters of these
Pad\'e approximants are known, it is possible to derive analytically
the subtracted VPF, $\widehat{\Pi}(\hat Q^2)$. With respect to an
approach where $\Pi(\hat Q^2)$ is directly used to determine
$\Delta\alpha_{\rm QED}^{\rm had}$, a benefit of the Adler function in
a global analysis of the lattice ensembles is that it requires
significantly fewer fit parameters. Indeed, for each lattice ensemble,
$\Pi(0)$ cancels in $D(\hat Q^2)$.

\begin{figure}[t!]
  \centering
  \subfigure[\label{fig:DQa}]{
    \includegraphics[height=0.39\linewidth]{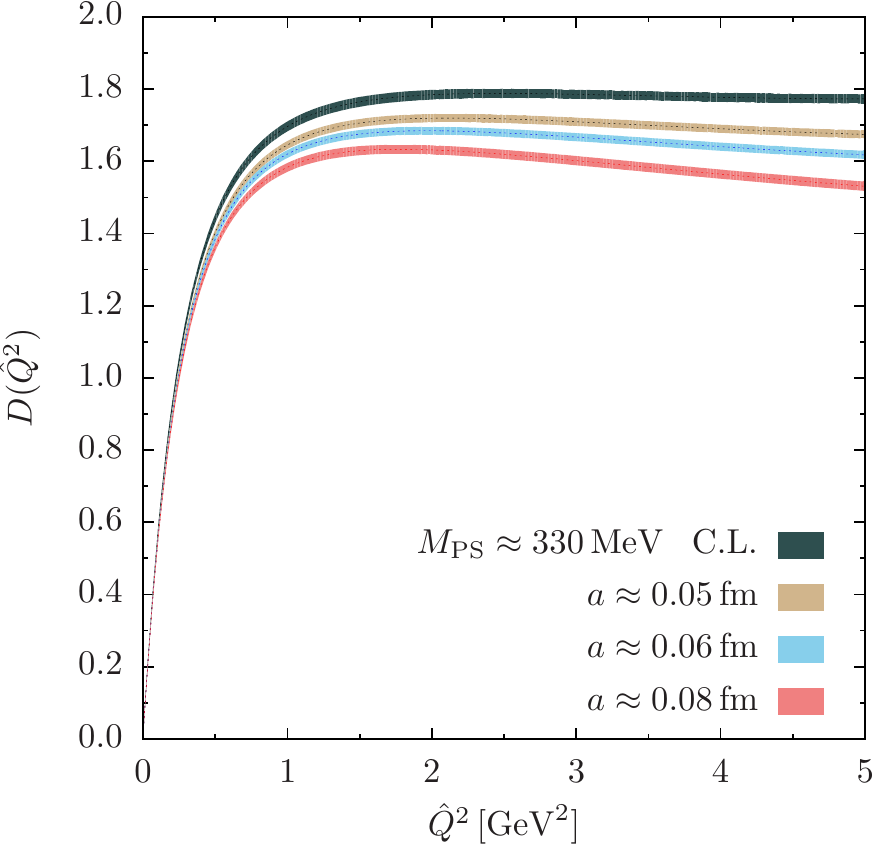}}
  \qquad
  \subfigure[\label{fig:DQMpi_b}]{
    \includegraphics[height=0.39\linewidth]{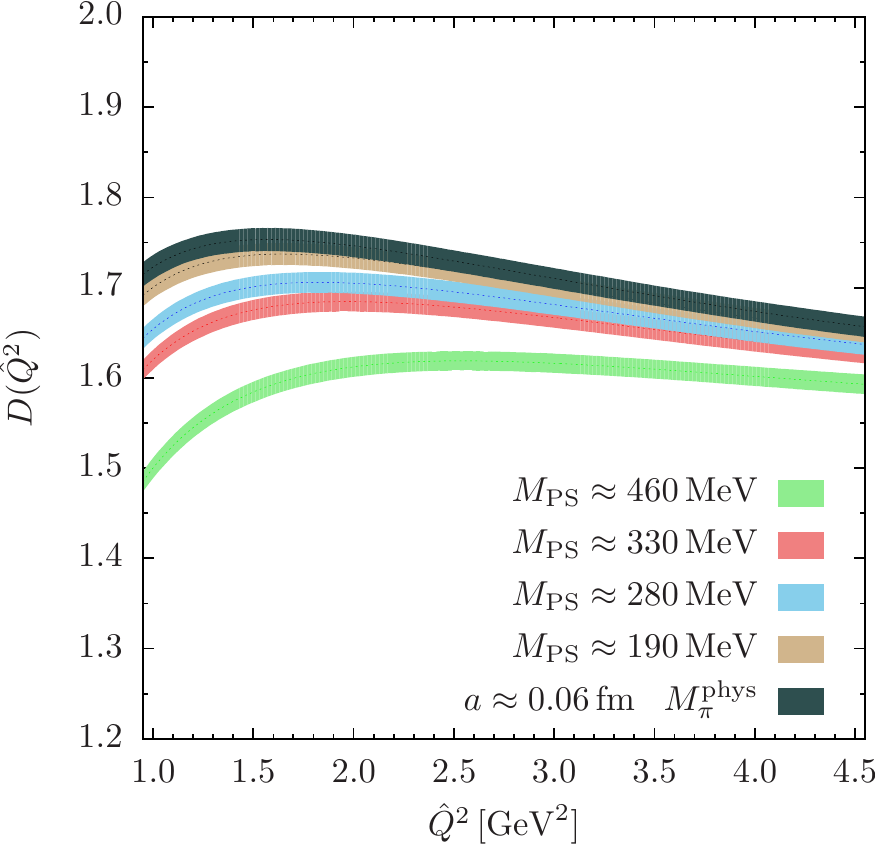}}
  \caption{Dependence on the momentum $\hat Q^2$ of the $(u,d)$
    contribution to the Adler function. (a) The coloured bands
    indicate the lattice spacing dependence of $D(\hat Q^2)$ at a fixed
    pion mass, $M_{\rm PS} \approx 330\,{\rm MeV}$. The dark upper
    band, labelled `C.L', is the continuum limit estimate. Cutoff
    effects are observed to grow for increasing values of $\hat Q^2$. (b)
    The effect of reducing the pion mass from $460\,{\rm MeV}$ down to
    $190\,{\rm MeV}$ is illustrated by the coloured bands. The dark
    upper band is the result of the extrapolation to the physical
    point. Light-quark mass effects are suppressed when increasing
    $\hat Q^2$.}
  \label{fig:DQ}
\end{figure}

An illustration of the pion mass dependence of the $(u,d)$
contribution to the Adler function -- at fixed $\hat Q^2$ -- is shown
in Fig.~\ref{fig:DMpi}. The coloured bands show the result of the
global fit of the ensembles in table~\ref{tab:ens}. In the momentum
region, $\hat Q^2 \geq 1\,{\rm GeV}^2$, that is most useful for the
determination of $\Delta\alpha_{\rm QED}^{\rm had}(\hat Q^2)$, we
observe that the lattice data can be well described by a linear
dependence on $M_{\rm PS}^2$. We have repeated the analysis by
excluding pion masses above $400\,{\rm MeV}$ to examine systematic
effects in the chiral extrapolation. The $\hat Q^2$ behaviour of the
$(u,d)$ contribution to the Adler function is shown in
Fig.~\ref{fig:DQ}. The size of lattice artefacts and of light quark
mass effects -- as a function of $\hat Q^2$ -- can be inspected
from the left and the right panels of Fig.~\ref{fig:DQ}, respectively.

\begin{figure}[t!]
  \centering
  \subfigure[\label{fig:DQNf}]{
    \includegraphics[height=0.39\linewidth]{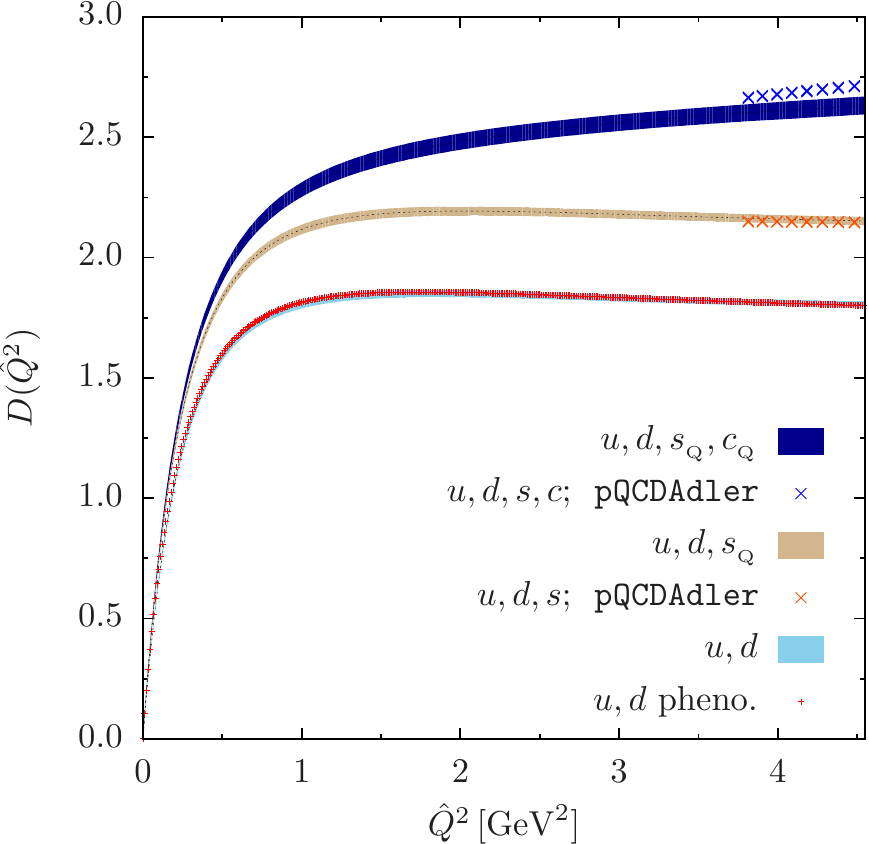}}
  \qquad
  \subfigure[\label{fig:alphaQNf}]{
    \includegraphics[height=0.39\linewidth]{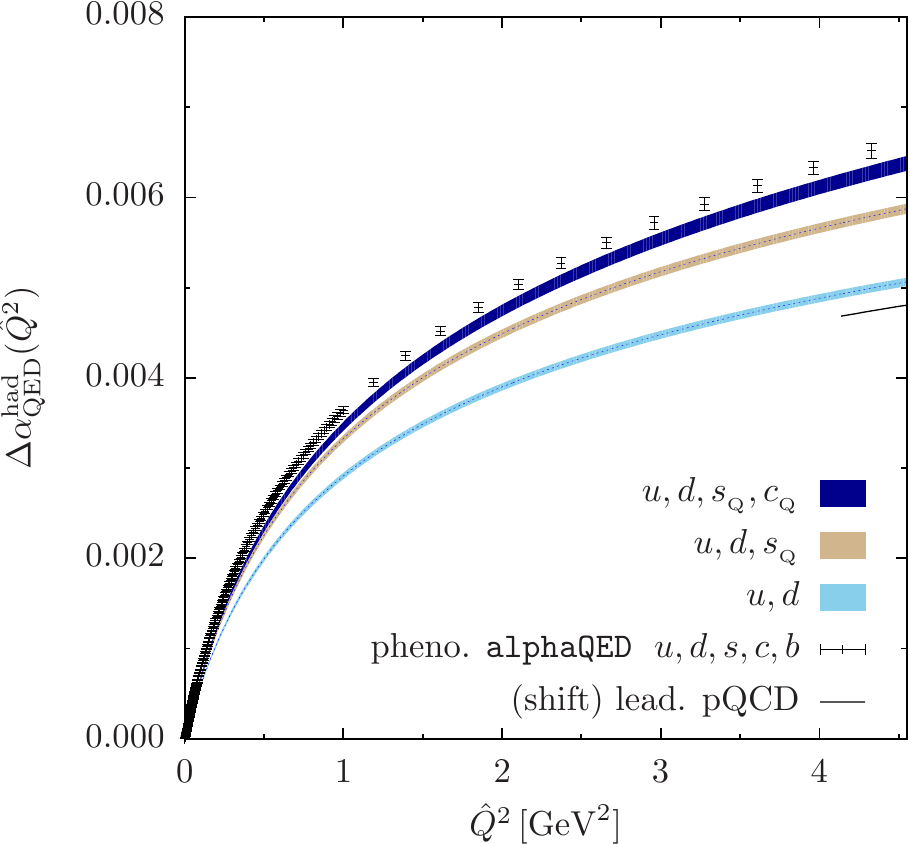}}
  \caption{(a) Contributions from $(u,d)$ and from partially quenched
    strange $s_Q$ and charm $c_Q$ quark flavours to $D(\hat Q^2)$
    after having performed the continuum and chiral
    extrapolations. The $(u,d)$ contribution is consistent with the
    phenomenological model of ref.~\cite{Bernecker:2011gh} denoted by
    the red `+' symbols. For the cases where $s_Q$ and $c_Q$ are
    included, perturbative QCD (pQCD) results from the
    \texttt{pQCDAdler} package~\cite{pqcdadler} are also shown. (b)
    Hadronic contribution to the running of the QED coupling for
    $(u,d)$, $s_Q$ and $c_Q$ quark flavours. The five-flavour result
    from the dispersion relation approach -- implemented in the
    package~\texttt{alphaQED}~\cite{Jegerlehner:2011mw,pqcdadler} --
    is shown by the black data points. The leading-order pQCD result
    for the $(u,d)$ case is shown by a small continuous black line
    that was shifted vertically to improve the visibility. The lattice
    error bands denote the size of statistical uncertainties only.}
  \label{fig:DalQ}
\end{figure}

Similar analyses were performed for the contributions to the Adler
function from partially quenched strange $s_Q$ and charm $c_Q$
quarks. The results from the various flavour contributions to $D(\hat
Q^2)$ in the continuum limit and at the physical point are presented
in Fig.~\ref{fig:DQNf}.
The Adler function shows a substantial dependence on $\hat Q^2$ at the
characteristic energy scale of QCD. For increasing values of $\hat
Q^2$, we observe a clear separation among the various flavour
contributions and the approach towards the pQCD behaviour.

Preliminary results for the running of the QED coupling are shown in
Fig.~\ref{fig:alphaQNf}. When increasing the flavour content up to the
inclusion of $(u,d)$, $s_Q$ and $c_Q$ contributions, the lattice
result approaches the five-flavour result from a phenomenological
analysis based on the dispersive approach. The bands in
Fig.~\ref{fig:alphaQNf} refer to lattice results in the continuum and
at the physical point with uncertainties that are purely
statistical. While a detailed comparison of lattice and
phenomenological results -- together with a complete account of
systematic effects -- is in progress, it is already interesting to
note that the statistical errors from the lattice are comparable to
those of the dispersive approach for $Q^2 \gtrsim 1\,{\rm
  GeV}^2$. This is contrast with the behaviour in the low-$Q^2$ region
-- which is most relevant for $a_\mu^{\rm HLO}$ -- where the lattice
data is affected by larger fluctuations. We note that differences and
ratios of $\Delta\alpha_{\rm QED}^{\rm had}$ between two different
scales can be built to further reduce the overall uncertainties. There
are, therefore, good prospects for an accurate determination of
$\Delta\alpha_{\rm QED}^{\rm had}$ from lattice QCD.


\section{Matching of the Adler Function to pQCD}

The Adler function is a useful quantity to monitor the regime of
momenta where pQCD is applicable. Non-perturbative effects can be
incorporated in the pQCD expansion through an operator product
expansion (OPE) where the operator matrix elements capture the
long-range strong interaction effects while the perturbative Wilson
coefficients encode the short distance physics. The matching of a
lattice determination of $D(\hat Q^2)$ to its OPE counterpart has to
be performed at large enough $\hat Q^2$ values to guarantee the
convergence of the pQCD expansion. However, lattice artefacts increase
with $\hat Q^2$ and should therefore fulfil the condition, $(a\hat
Q)^2 \ll 1$, to be kept under control. A first account of our
investigations about the possibility to determine $\Lambda_{\rm QCD}$
from a matching to pQCD of the VPF evaluated on the lattice was
reported in~\cite{Herdoiza:2014jta}.

Non-singlet contributions to $D(Q^2)$ are considered both on the
lattice and the pQCD sides. The OPE of the Adler function reads,
\begin{eqnarray}
  {D_{\rm OPE}}(Q^2,\alpha_s,m_{\rm f}) \, &=&\, D_0(\alpha_s,Q^2,\mu^2)
  +\, D_2^m(\alpha_s,Q^2,\mu^2)\,
  \frac{\left(\mbar_{\rm f}[Q^2]\right)^2}{{Q^2}}  +\, D_4^{\rm
    F}(\alpha_s,Q^2,\mu^2)\,\frac{\mbar_{\rm f} \langle \bar \psi_{\rm f}
    \psi_{\rm f}\rangle}{{Q^4}} \nonumber\\ &+&\, D_4^{\rm
    G}(\alpha_s,Q^2,\mu^2)\,\frac{\langle O^{(4)}_{\rm OPE}
    \rangle}{{Q^4}} + \, \mathcal{O}\left(\frac{1}{{Q^6}}\right)\,,
  \label{eq:Dope}
\end{eqnarray}
where the flavour content, ${\rm f}=(u,d)$, is considered. The Wilson
coefficients $D_0$, $D_2^m$, $D_4^{\rm F}$ and $D_4^{\rm G}$ in
eq.~(\ref{eq:Dope}) are computed in pQCD and, depending on the case,
are known from 1- to 4-loop order in the $\alpha_s$
expansion~\cite{Chetyrkin:1996cf,Baikov:2012zm,Chetyrkin:1985kn}. Non-perturbative
effects are encoded through the condensates $\langle \psibar_{\rm f}
\psi_{\rm f}\rangle$ and $\langle O^{(4)}_{\rm OPE} \rangle$. The pQCD
expressions are defined in the $\overline{\mathrm{MS}}$ scheme. The
connection of the coupling ${\alpha_s}$ to the scale
${\Lambda_{\overline{MS}}^{(\nft)}}$ is given by the 4-loop
$\beta$-function~\cite{vanRitbergen:1997va}.  The renormalised quark
mass $\mbar_{\rm f}(\mu^2)$ is given at the renormalisation scale
$\mu$~\cite{Fritzsch:2012wq,Chetyrkin:1999pq}. Since the Adler
function is a physical quantity, any residual scheme and scale
dependence should vanish as higher order terms are included.

\begin{figure}[t!]
  \centering
  \subfigure[\label{fig:DQope}]{
    \includegraphics[height=0.39\linewidth]{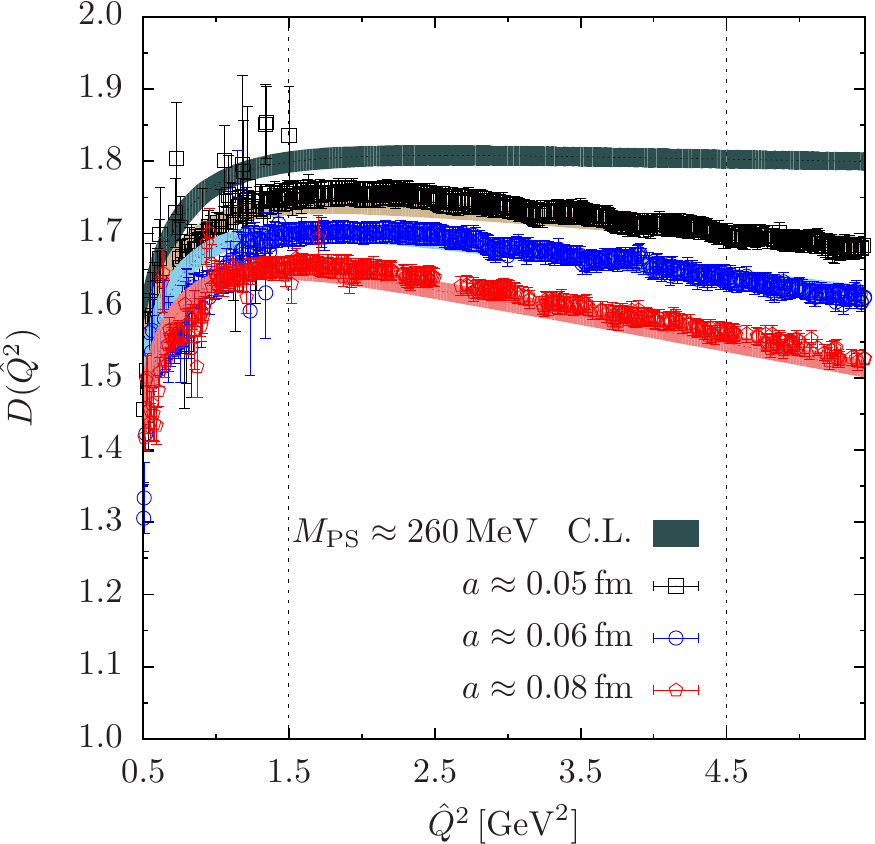}}
  \qquad
  \subfigure[\label{fig:DQopepade}]{
    \includegraphics[height=0.39\linewidth]{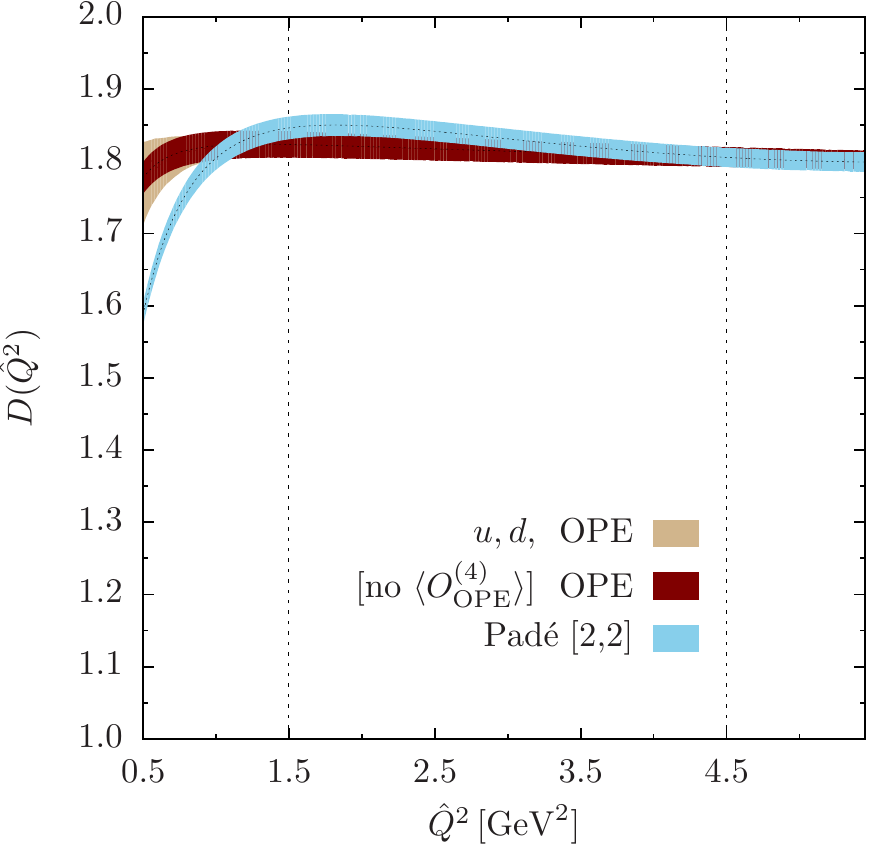}}
  \caption{Dependence on the momentum $\hat Q^2$ of the $(u,d)$
    contribution to the Adler function. (a) The lattice data from
    ensembles at three different values of the lattice spacing and
    $M_{\rm PS} \approx 260\,{\rm MeV}$ are fitted to an OPE
    expression in eq.~(\protect\ref{eq:Dope}) supplemented with terms
    parametrising lattice artefacts. The dark upper band, labelled
    `C.L', is the continuum limit estimate. (b) Comparison of the use
    of OPE and Pad\'e ans\"atze in the determination of the Adler
    function after having performed the continuum and chiral
    extrapolations. Signs of compatibility are observed only at
    sufficiently large $\hat Q^2$ values, where the perturbative
    series is better behaved. The effect of removing from the OPE the
    contribution for the dimension-four condensate is also shown. In
    both panels, the $\hat Q^2$-interval considered in the OPE fit is
    denoted by the vertical dashed lines and the bands denote
    statistical errors only.}
  \label{fig:Dope}
\end{figure}

The measurements of $D(\hat Q^2)$ from the complete set of ensembles
in table~\ref{tab:ens} are matched to the OPE expression in
eq.~(\ref{eq:Dope}) where $\alpha_s(\mu=2\,{\rm GeV})$ and $\langle
O^{(4)}_{\rm OPE}\rangle$ are left as fit parameters. The fit ansatz
is augmented by two terms parametrising $\hat Q^2$-dependent and
-independent discretisation effects, respectively. For the RGI product, $\mbar_{\rm
  f}\, \langle \psibar_{\rm f} \psi_{\rm f}\rangle$, we use as input
the value of the chiral condensate, $\langle \psibar_{u}
\psi_{u}\rangle = -(0.269(8)\,{\rm GeV}^3)$~\cite{Aoki:2013ldr}.

Fig.~\ref{fig:DQope} shows an example of an OPE fit from $1.5-4.5\,{\rm GeV}^2$
that confirms that a proper control of lattice artefacts is essential
for the matching to pQCD. The continuum estimate of $D(\hat Q^2)$
based on Pad\'e approximants (c.f.~sect.~\ref{sec:alp}) can be
compared to the results of the OPE fit. This is illustrated in
Fig.~\ref{fig:DQopepade}, where the effect of removing the $\langle
O^{(4)}_{\rm OPE} \rangle$ term in eq.~(\ref{eq:Dope}) is also
shown. In the $\hat Q^2$-interval, $1.5-4.5\,{\rm GeV}^2$, where
the OPE fit is performed, deviations at the two-sigma level can be
observed -- based on statistical errors only. The expected breakdown
of the perturbative expansion is manifest for lower $\hat Q^2$
values. Barring yet uncontrolled systematic effects,
Fig.~\ref{fig:DQopepade} suggests that momenta, $Q^2 \gtrsim 4\,{\rm
  GeV}^2$, might be needed to observe reliable signs of convergence of the pQCD
expansion of the Adler function.


\section*{Conclusions}

The electromagnetic coupling is an input parameter in a number of
precision studies of the Standard Model. An accurate determination of
its running is essential to constrain many of these processes. Our
studies provide evidences for the potential advantage of a lattice
calculation of the Adler function $D(Q^2)$ to achieve a rather precise
determination $\Delta \alpha_{\rm QED}^{\rm had}(Q^2)$. Furthermore, a
comparison of $D(Q^2)$ to an approximation based on the OPE suggests
that momenta, $Q^2 \gtrsim 4\,{\rm GeV}^2$, are needed to observe
reliable signs of convergence of the perturbative expansion. Ongoing
studies will provide a more complete account of the systematic effects
present in these determinations.


\paragraph*{Acknowledgements}

Our calculations were performed on the ``Wilson'' and ``Clover'' HPC
Clusters at the Institute of Nuclear Physics, University of Mainz.
We thank Dalibor Djukanovic and Christian Seiwerth for technical support.
This work was granted access to the HPC resources of the Gauss Center
for Supercomputing at Forschungzentrum J\"ulich, Germany, made available
within the Distributed European Computing Initiative by the PRACE-2IP,
receiving funding from the European Community's Seventh Framework
Programme (FP7/2007-2013) under grant agreement RI-283493 (project
PRA039).
We are grateful for computer time allocated to project HMZ21 on the
BG/Q JUQUEEN computer at NIC, J\"ulich.
This research has been supported by the DFG in the SFB~1044.
We thank our colleagues from the CLS initiative for sharing
the ensembles used in this work.
G.H. acknowledges support by the Spanish MINECO through the Ram\'on y
Cajal Programme and through the project FPA2012-31686 and by the
Centro de Excelencia Severo Ochoa Program SEV-2012-0249.


\end{document}